\documentclass[twocolumn,abs,prb,longbibliography]{revtex4-1}
\usepackage{graphicx}
\usepackage{dcolumn}
\usepackage{float}
\usepackage{amsmath}
\usepackage{bm}

\usepackage{booktabs}
\newcolumntype{.}[1]{D{.}{.}{#1}}
\usepackage{color}
\usepackage{hyperref}
\hypersetup{
	breaklinks=true,
	pdfnewwindow=true,      
	colorlinks=true,       
	linkcolor=blue,          
	citecolor=blue,        
	filecolor=blue,      
	urlcolor=blue,          
}
\begin{document}
\title{Optical properties of superconducting pressurized LaH$_{10}$}
\author{S. F. Elatresh}
\email[Electronic address:]{sabri.elatresh@dal.ca,enicol@uoguelph.ca}
\affiliation{Department of Physics, University of Guelph, Guelph, Ontario, Canada N1G 2W1}
\affiliation{Department of Chemistry and Chemical Biology, Cornell University, Baker Laboratory, Ithaca, New York 14853-1301, USA}
\author{T. Timusk}
\affiliation{Department of Physics and Astronomy, McMaster University, Hamilton, Ontario L8S 4M1, Canada}
\author{E. J. Nicol}
\email[Electronic address:]{sabri.elatresh@dal.ca,enicol@uoguelph.ca}
\affiliation{Department of Physics, University of Guelph, Guelph, Ontario, Canada N1G 2W1}

\date{\today}

\begin{abstract}
Recently superconductivity has been discovered at around 200~K in a hydrogen sulfide system and around 260~K in a lanthanum hydride system, both under
pressures of about 200 GPa. These record-breaking transition temperatures bring within reach the long-term goal of obtaining room temperature superconductivity. We have used first-principle calculations
 based on density functional theory (DFT) along with Migdal-Eliashberg theory to investigate the electron-phonon mechanism for superconductivity in the $Fm\bar{3}m$ phase 
 proposed for the LaH$_{10}$ superconductor. We show that the very high transition temperature  $T_c$ results from a highly optimized electron-phonon interaction that favors coupling to high frequency hydrogen phonons. Various superconducting properties are calculated, such as the energy gap, the isotope effect, the specific heat jump at $T_c$, the thermodynamic critical field and the temperature-dependent penetration depth. However, our main emphasis is on the finite frequency optical properties, measurement of which may allow for an independent determination of $T_c$ and also a confirmation of the mechanism for superconductivity. 
\end{abstract}

\date{\today}
\pacs{61.50.Ks,62.50.+p}
\maketitle

\section{Introduction}
Neil Ashcroft proposed that hydrogen, upon becoming metallic at sufficiently high density, 
would transform to a superconductor at a critical transition temperature $T_c$ possibly well above room temperature~\cite {PhysRevLett.21.1748}. 
This stimulated many scientists to search for superconductivity not only
 in pure hydrogen but also in hydrogen-rich compounds. The possible formation of metallic
  hydrogen has been reported at very high pressure by a number of groups~\cite {Mao:1989,Eremets:2011,Dias715,Loubeyre:2020}. Moreover, ab-{\it initio} crystal structure calculations 
 have predicted high-temperature superconductivity in some hydrogen-rich 
 compounds, some of which have been confirmed experimentally~\cite {floreslivas2019perspective,Zhang2017,Oganov2019,Pickard:2020}. 

 Among a large number of possible hydrogen-rich compounds, several stoichiometries for the H-S system were identified as energetically competitive~\cite{PhysRevB.93.020103} and experimentalists had been looking for superconductivity by compressing H$_{2}$S with H$_{2}$ ~\cite {PhysRevLett.107.255503}. H$_{2}$S is the only known stable compound in the system at ambient pressure and room temperature and its highest $T_c$ was found to be 80~K at 160~GPa ~\cite{doi:10.1063/1.4874158}.
 In 2014, Duan {\it et al.}~\cite{Duan2014} theoretically predicted that H$_{3}$S was a good candidate for superconductivity in the H-S system, with an expected  $T_c$   as high as 191-204 K at 200~GPa. This was followed by a remarkable discovery by
Drozdov {\it et al.}~\cite{Drozdov2015} who observed superconductivity at 203 K in H$_{2}$S 
compressed to 155~GPa (now identified as transforming to H$_{3}$S), which 
set the record for the highest $T_c$ at the time.  
This subsequent experimental discovery helped to establish DFT calculations as having predictive power for finding new high temperature superconductors in the pressurized hydrides. 
Theoretical  structure calculations and experimental studies have since adopted several structures for H$_3$S, including  the $R3m$, $Im\bar{3}m$, and $Cccm$ phases~\cite {Duan2014,PhysRevB.91.180502, PhysRevLett.114.157004,PhysRevB.93.020103,PhysRevB.91.060511,Errea2016,Einaga2016,PhysRevB.95.020104,PhysRevB.93.174105}, but the interpretation of X-ray data has favored the $Im\bar 3m$ phase for the high pressures involved.

 \begin{figure} [t!]
 \centering
\includegraphics[width=0.44\textwidth, clip]{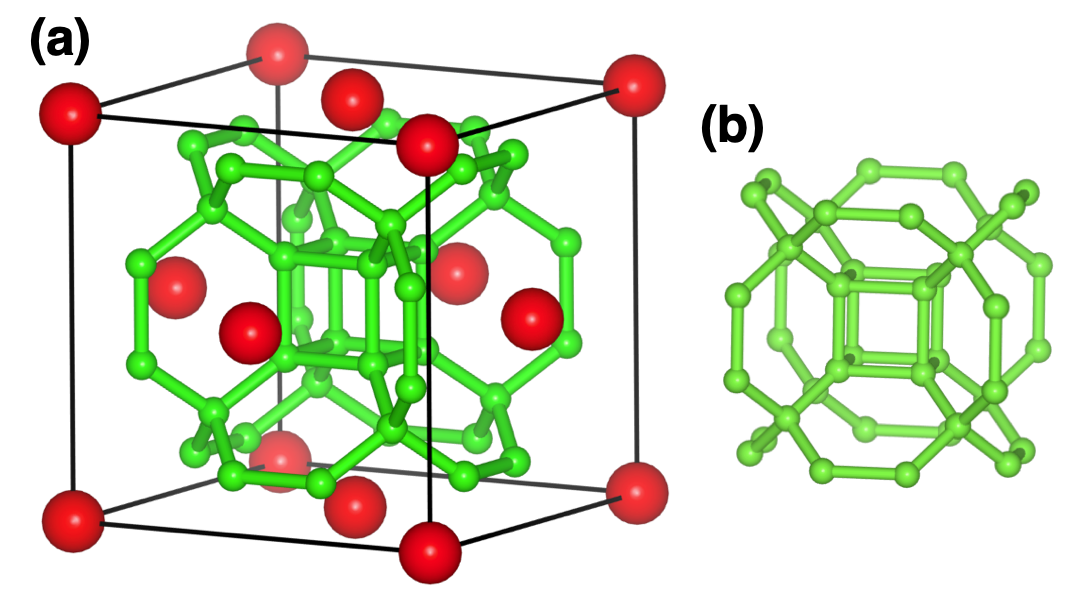}
\caption{(a) The conventional unit cell of the $Fm\bar3 m$ crystal structure of LaH$_{10}$ where the red balls represent the La atoms in a fcc structure and  
the green balls are the hydrogen atoms. (b) The hydrogen atom arrangement shown separately. \label{LaH10} }
\end{figure}

The record-breaking discovery of very high $T_c$ in H$_3$S has encouraged further efforts in the search for 
 room temperature superconductivity in pressurized hydride materials. Several superconducting hydride materials have been 
 predicted, including MgH$_{6}$ with $T_c$=271~K at 300~GPa~\cite{C5RA11459D}, CaH$_{6}$ with  $T_c$=~235~K at 150~GPa~\cite{Wang6463}, 
 and YH$_{6}$ with $T_c$=~264~K at 120~GPa~\cite {Li2015}. Interestingly, two independent first-principles structure search studies revealed some 
 stable H-rich clathrate structures to be potential room-temperature superconductors~\cite{Liu6990,PhysRevLett.119.107001}. In particular, the La-H and Y-H systems were
  theoretically investigated, and some compounds have been identified as good candidates for room-temperature superconductivity. From, the Y-H system, the 
  structures  YH$_{4}$,  YH$_{6}$, and  YH$_{10}$ were proposed~\cite{Li2015, Liu6990, PhysRevLett.119.107001}. While from the La-H system several stable compounds were
   identified: LaH$_{3}$, LaH$_{4}$, LaH$_{5}$, LaH$_{8}$, and LaH$_{10}$.  Most notably, YH$_{10}$ and LaH$_{10}$ were considered  the most likely structures 
   to exhibit possible room-temperature superconductivity, where the $T_c$ $\sim$~303~K at 400~GPa~\cite{PhysRevLett.119.107001} for 
    YH$_{10}$ and $T_c$ $\sim$ 274-286~K at 210~GPa for the $Fm\bar{3}m$ phase
    proposed for the LaH$_{10}$. As a result, the La-H system became of interest, with studies proposing LaH$_{10}$ to be thermodynamically stable 
    above 150~GPa. Indeed, the lanthanum compound was  synthesized~\cite {doi:10.1002/anie.201709970} and two groups have reported near-to-room-temperature
     superconductivity with  $T_c$ $\sim$ 260~K at 190~GPa~\cite{PhysRevLett.122.027001} and $T_c$ $\sim$~250~K at about 170~GPa~\cite{Drozdov2019}. 
     This is the highest $T_c$ that has been reported to date.  Various structures and compositions have been suggested for
      the superconducting La-H system  including $P6/mmm$ for  LaH$_{16}$~\cite {PhysRevB.101.024508} and $R\bar{3}m$, $C2/m$,  $Fm\bar{3}m$ for the LaH$_{10}$ 
      composition. However, recently Errea, {\it et al.}~\cite{Errea2020} have shown that the Born-Oppenheimer energy 
surface with classical treatment  displays many local minima and  by including quantum effects all these phases collapse to a single phase, $Fm\bar3m$, which is highly 
 symmetric.

In this work, we report results from first-principles calculations based 
on DFT and Migdal-Eliashberg theory, to 
evaluate $T_c$ and other superconducting properties of  LaH$_{10}$. In particular, we focus on studying the
 most promising candidate structure proposed for LaH$_{10}$, the $Fm\bar{3}m$ phase. We emphasize the finite frequency optical conductivity which is related to reflectance. The latter is a technique accessible to experiment in high pressure cells and has been used with success in H$_3$S~\cite{Capitani2017}. It can be used to identify important fundamental information about superconductivity such as the binding energy of electrons in the superconducting condensate and the mechanism for binding, which for conventional superconductors is the electron-phonon interaction. Moreover, it has been proposed that it can provide an independent measurement of $T_c$ for the pressurized hydrides~~\cite{PhysRevLett.121.047002}.

 The structure of our paper is as follows. 
 In Sec.~II, we begin by calculating the electronic band structure and density of states, illustrating the Fermi surface. This is followed by a calculation of the phonon dispersion curves and the phonon density of states in the harmonic approximation. These are then used further to calculate the electron-phonon interaction shown in Sec.~III. Proceeding to the calculation of superconducting properties in Sec.~III, we calculate $T_c$, the functional derivative, the isotope effect, and the important BCS ratios for the energy gap, the specific heat jump at $T_c$ and the thermodynamic critical field. We then proceed to the calculation of the temperature-dependent superfluid density (or inverse square of the penetration depth) followed by the finite frequency  optical conductivity and reflectance. To examine the influence of anharmonicity on these properties, we also provide comparison calculations based on an anharmonic spectrum taken from Errea {\it et al.}~\cite{Errea2020}. We provide our final summary in Sec.~IV.

\section{Electronic and Phononic Properties}

 \begin{figure} [t!]
 \centering
\includegraphics[width=0.47\textwidth, clip]{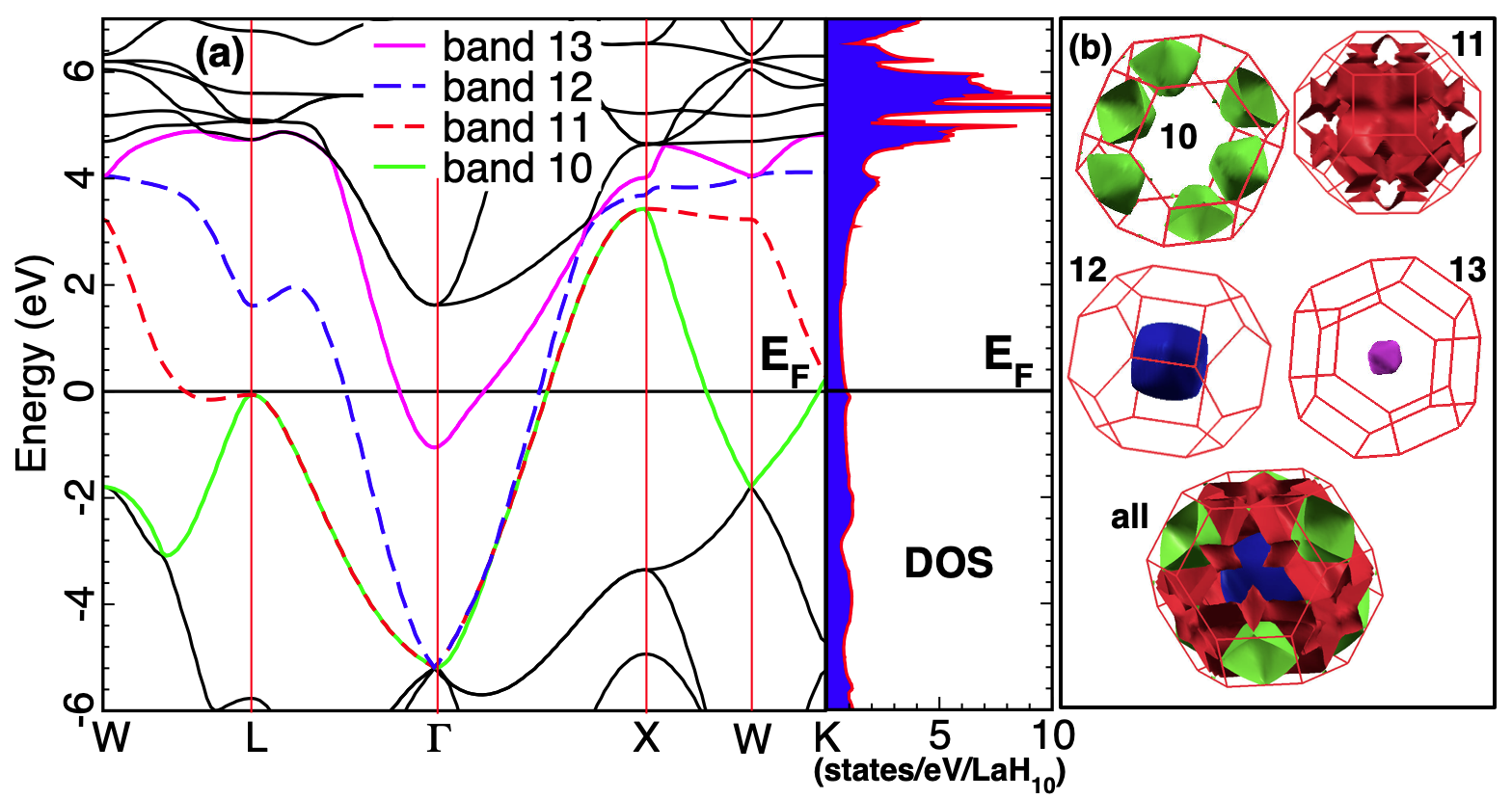}
\caption{\label{eband_FS}{Calculated electronic properties for LaH$_{10}$ at 210~GPa: (a) the electronic 
band structure and the density of states (DOS) and (b) the Fermi surface. Each band crossing the Fermi level $E_F$ in (a) is given a color which corresponds with the piece of Fermi surface with the same color in (b). }}
\end{figure}

We begin by addressing the electronic and the phononic properties, as well as, the electron-phonon coupling. It has been suggested that the LaH$_{10}$ $Fm\bar3m$ phase has a wide pressure range for dynamic stability,
down to about 210~GPa  for harmonic phonon calculations~\cite{Liu6990} and lower for anharmonic quantum calculations~\cite{Errea2020}, in the range where superconductivity has been observed. Consequently, we have calculated 
the electronic band structure and density of states for LaH$_{10}$ at 210~GPa (Fig.~\ref{eband_FS}(a)).
Full structural optimization, electronic band structure, density of states,  and
Fermi surfaces  calculations
were performed within density-functional theory (DFT)  
using the  ABINIT code~\cite{Gonze20092582} with 1 and 11-electron
 Hartwigsen-Goedeker-Hutter pseudopotential~\cite{PhysRevB.58.3641} for H and La, 
respectively, and the local density approximation (LDA) by Teter Pade parameterization 
~\cite {PhysRevB.54.1703}.  A plane-wave expansion with 50 Ha cut-off and a {\bf k}-point 
grid for self-consistent calculations as large $24\times24\times24$. These
  dense {\bf k}-point grids  are sufficient to ensure convergence for enthalpies to better than 1~meV/atom.
The symmetry information of the LaH$_{10}$ $Fm\bar3m$ phase and the calculated X-ray diffraction obtained after performing full structural optimization at 210 GPa can be found in Table~S1 and Fig.~S1~\cite{Supplemental}, respectively.
  The electronic band structure indicates that LaH$_{10}$ at 210 GPa is a good metal, which is consistent with previous calculations 
  reported at 300~GPa~\cite{Liu6990}. As was expected, the partial electronic density of states (Fig.~S2~\cite{Supplemental}) indicates that the La contribution is greater than H.
  There are four bands crossing the
    Fermi level and the Fermi surface is shown in Fig.~\ref{eband_FS}(b) in colors which correspond to each of the four bands.
    The Fermi surfaces of bands 10, 12, and 13 are quite simple while the Fermi surface of band 11 is quite complicated. Similar pictures are also found in Ref.~\cite{Wang:2020} which discusses the possibility of multiband superconductivity associated with different Fermi surfaces.

    The density of states at the Fermi level, N($E_F$), has a large number of states, 11.85 states/Ry, which
 is larger by factors of 1.18 and 1.58  than fcc LaH$_{10}$  and H$_3$S reported at 
 300~GPa~\cite {Liu6990} and 200~GPa ~\cite{PhysRevB.91.060511}, respectively. This variation with pressure is consistent with the recent 
 finding by Errea et al.~\cite{Errea2020} that the density of states
  at the Fermi level decreases with increasing pressure.  A close up view of our calculation of the electronic density of 
  states near the Fermi level (Fig. S3~\cite{Supplemental}) shows a peak in the density of states just below $E_F$, which likely has its primary origin in the 
  flattening of band 11 near the L point. A large density of states at the Fermi level is favorable for superconductivity.

Now we discuss the phonon characteristics, as they play an important role in electron-phonon mediated superconductivity.  We have performed 
 phonon calculations for the $Fm\bar{3}m$ phase for LaH$_{10}$ at 210~GPa, in a pressure region where 
 measurements confirm a very high $T_c$.
The phonon and the electron-phonon coupling (EPC) calculations were performed with density-functional  perturbation 
theory (DFPT)~\cite{PhysRevB.55.10337} and the harmonic approximation as implemented in ABINIT code~\cite{Gonze20092582} with the same parameters as in the DFT calculations.
The dynamical matrices were calculated initially on
uniform {\bf q}-point meshes, from which interatomic force constants
are obtained and used to interpolate the phonon dispersions over the
entire Brillouin zone. The phonon band structure and density of states for the
{\bf q}-point grid was tested
until convergence was achieved. It was found that a $6\times6\times6$ grid was required. The EPC implementation in the
 ABINIT code~\cite{Gonze20092582} is similar to  that implemented by 
 Liu  and Quong~\cite {PhysRevB.53.R7575} and Savrasov~\cite {PhysRevB.54.16487}.

 \begin{figure} [t!]
 \centering
\includegraphics[width=0.43\textwidth, clip]{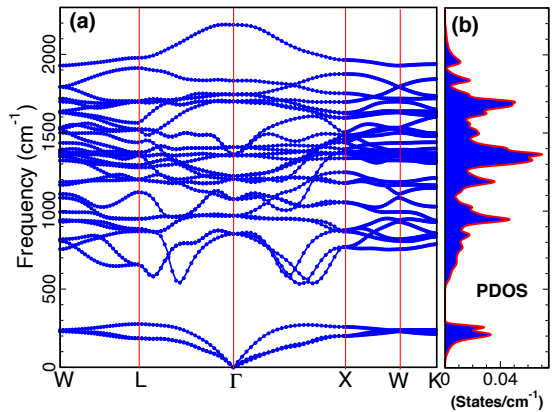}
\caption{\label{PDOS} Calculated phonon properties for LaH$_{10}$ at 210~GPa: (a) the phonon dispersion curves and (b) the phonon density of states (PDOS).}
\end{figure}

 The phonon band structure and density of states for the
 LaH$_{10}$-$Fm\bar{3}m$ phase is shown in Fig.~\ref{PDOS} and  they 
 are similar to previous calculations at 300~GPa~\cite{Liu6990} where GGA was used. This highlights that both GGA and LDA provide a consistent description of vibrational properties of the LaH$_{10}$-$Fm\bar{3}m$ phase. It is clear from this that the
  phonon frequencies have a weak dependence on increasing pressure, which indirectly might imply a weak dependence of $T_c$ on pressure ~\cite{Drozdov2019,Errea2020}. 
  The highest phonon frequency for LaH$_{10}$ at 210~GPa is at $\sim$ 2190~cm$^{-1}$ or $\sim 272$ meV (Fig.~S4~\cite{Supplemental}). The very lowest phonon branches result from the heavy mass lanthanum phonon modes and the rest of the dispersion curves at higher frequencies are due to the lighter mass hydrogen. The role of the lanthanum appears to be that of providing conduction electrons and also caging the hydrogen to provide a shorter H bond length than normal. The La-La bond length in the fcc structure underlying LaH$_{10}$ (Table S1, Figs. S5 and S6~\cite{Supplemental}) is typical of ordinary pure metals at ambient pressure but the shortened distance between the hydrogen is not. Rather, the
  shortest H-H distance in LaH$_{10}$ at 210~GPa (Figs. S5 and S7~\cite{Supplemental}) is 1.084 \AA, which is more consistent with  the
  prediction of Neil Ashcroft  for metallic hydrogen at a pressure near 500~GPa~\cite {PhysRevLett.21.1748}. 
  The shortest La-H distance is 2 \AA (Fig. S8~\cite{Supplemental}). As a result, the La cage might be viewed as assisting in shortening the H-H distance so that the 500~GPa needed for  pure metallic hydrogen is now only of order 200~GPa for metallic behavior.

  Finally, we note that our calculation of the harmonic phonon dispersion curves at 210~GPa does not show any unstable modes. Interestingly, Errea et al.~\cite {Errea2020} have reported that unstable
modes appear in the harmonic approximation at pressures below $\sim$ 230~GPa and that including anharmonicity
 is required to stabilize this structure below 230~GPa. We have investigated this matter further. The pressure where the 
 harmonic approximation breaks down may be varied depending on the choice of some DFT parameters, including the type 
 of exchange-correlation and pseudopotential~\cite{PhysRevB.97.214101}.  We have produced a result with unstable modes when we
  used harmonic calculation parameters similar to Ref.~\cite {Errea2020}. To examine the quality of the harmonic electron-phonon spectral function used in this work and the influence 
  of anharmonicity, we have included comparison calculations, where an anharmonic  
  electron-phonon spectrum from Errea et al.~\cite {Errea2020} is used as input. 

  \section{Superconducting Properties}

\begin{figure} [t!]
 \centering
\includegraphics[width=0.44\textwidth, clip]{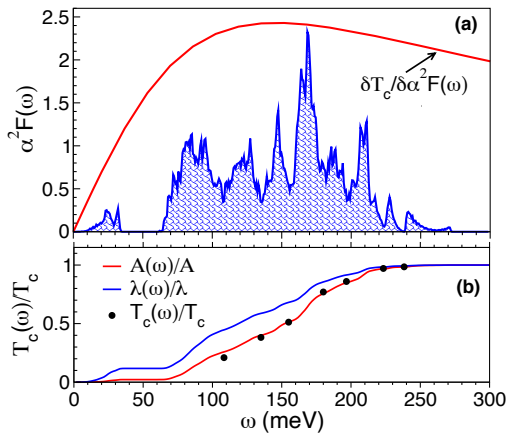}
\caption{(a) $\alpha^2F(\omega)$ for  the $Fm\bar{3}m$ phase of LaH$_{10}$ at 210~GPa.
  Also, shown is the functional derivative $\delta T_c/\delta\alpha^2F(\omega)$  as a function of 
  energy $\omega$ which indicates which of these energies are more effective for raising $T_c$. This curve has been scaled up by a factor of 15. (b)  $T_c(\omega)/T_c$ versus $\omega$
  as well as electron-phonon renormalization parameter $\lambda(\omega)/\lambda$ and area $A(\omega)/A$.\label{A2F}}
\end{figure}

We now continue on to the electron-phonon coupling and emphasize its contribution to superconducting properties. We have used our calculation of the electronic structure and phonon dispersion curves to calculate the electron-phonon spectral function $\alpha^2F(\omega)$ which enters the Migdal-Eliashberg theory of superconductivity and is the pairing ``glue'' for Cooper pairs in conventional superconductors~\cite{Carbotte:1990}. The $\alpha^2F(\omega)$ is shown in Fig.~\ref{A2F}(a) where we see that it looks much like the phonon density of states from Fig.~\ref{PDOS}(b) but with modified strength as a function of energy due the variation of the electron-phonon coupling vertex.

In the following, this $\alpha^2F(\omega)$ was used in our calculations of superconducting properties which are based on numerical iteration of the nonlinear coupled Migdal-Eliashberg equations for the real-frequency complex gap function $\Delta(\omega,T)$ and complex renormalization function $Z(\omega,T)$.~\cite{Carbotte:1990,Marsiglio:2008,Carbotte:2019} The linearized imaginary-frequency axis equations based on Matsubara frequencies $\omega_n$ were used to calculate $T_c$. Then the coupled nonlinear imaginary-axis equations were used to calculate thermodynamic properties for temperatures below $T_c$ ~\cite{Carbotte:1990,Marsiglio:2008}. An analytic continuation procedure was then used to evaluate the real frequency gap and renormalization functions from the imaginary axis solutions~\cite{Marsiglio:1988}. The real frequency axis quantities then allowed for the energy gap and other transport properties to be obtained. The finite frequency conductivity was evaluated based on formulas in Ref.~\cite{PhysRevB.43.12804,Marsiglio:2008}   and the reflectance was calculated following the prescription given in Ref.~\cite{Capitani2017}.
The formalism is lengthy to write down and has been presented in the literature numerous times, consequently, for formulas and further descriptions see Refs.~\cite{Carbotte:1990,Marsiglio:2008,Carbotte:2019}. We will emphasize results and discussion here.

 \subsection{The critical temperature}
We begin by discussing the characteristics of this spectral function
    and the features which have an impact 
    on increasing the $T_c$.
    Using the $\alpha^2F(\omega)$ spectrum in the linearized form of the Migdal-Eliashberg equations  
 on the imaginary frequency axis~\cite{Carbotte:1990,Carbotte:2019} and iterating these equations, we may obtain $T_c$ for a particular choice of
 Coulomb pseudopotential parameter $\mu^*$ or, alternatively, we can fix the value of $T_c$ and iterate
  for $\mu^*$ ~\cite{Carbotte:1990}. Historically, it has been typical to chose the $\mu^*$ to give the experimental $T_c$. Once $\mu^*$ is fixed, then 
  all other superconducting properties follow with no free parameters. For a typical value of $\mu^*=0.128$, we match the $T_c=237.9$~K that
   is found for an anharmonic calculation of $\alpha^2F(\omega)$ including quantum corrections at a similar pressure~\cite{Errea2020} [designated as the Errea spectrum]. This is also close to the 
   experimental $T_c$ for this pressure. This allows us to make comparison with an alternative calculation of $\alpha^2F(\omega)$ as we calculate 
   the superconducting properties of LaH$_{10}$. Indeed, in Table 1, we list the comparison of results from our spectrum with those that we have 
   calculated from the Errea spectrum.

\begin{table}[!h]
\centering
\caption{\rm Summary of quantities calculated using two independent $\alpha^2F(\omega)$ spectra and   fixing the $T_{c}$ = 237.9~K.}
 \resizebox{0.75\columnwidth}{!}{%
\begin{tabular}{l c c }
\hline  \hline   
{         } & {\it Errea spectrum} &  {\it  Our spectrum} \\
{         } & {\it (anharmonic)} &  {\it  (harmonic)} \\
{         } & {\it            } (214~GPa) &  {\it       }(210~GPa)\\
\hline
$\mu^{*}$ & 0.15 &  0.13  \\
$\lambda$  & 2.02 &  2.16  \\
$A$ (meV)       & 133 & 130\\
$\omega_{\rm ln}$ (meV)  &  113 & 102  \\
$T_c/A$  & 0.155 & 0.157  \\
$T_c/\omega_{\rm ln}$  & 0.18 & 0.20  \\
Isotope coeff. $\beta$ & 0.484 &  0.489  \\
$\Delta_0$ (meV)  & 49  & 51   \\
\hline
$2\Delta_0/k_BT_c$ & 4.78 &  4.93  \\
$\Delta C(T)/\gamma T_c$ & 2.97 &  2.91  \\
$\gamma T_c^2/H_c^2(0)$ & 0.131 &  0.130  \\
\hline \hline
\end{tabular}
}
\end{table}

   In Fig.~\ref{A2F}(a),
   we show the functional derivative of $T_c$ with respect to $\alpha^2F(\omega)$, this quantity indicates which
 frequencies in the spectrum are more important for increasing $T_c$. While the curve is entirely positive, indicating that adding 
 more weight in the spectrum at any frequency will increase $T_c$, it is clear that the broad maximum in this
  spectrum favors the region where we find the hydrogen phonons and that the low frequency La phonons have very little 
  effect on the $T_c$, indeed our calculations show that they only contribute a few degrees to $T_c$. Consequently, 
  the large $T_c$ is due to the electrons coupling to hydrogen phonons, as expected from simple arguments that $T_c$ should scale with phonon frequency. Similar results have been recently analyzed by others, for example Refs.~\cite{Quan:2019,Papa:2020}.

  Isotopic substitution is a classic experiment to confirm the electron-phonon interaction as the underlying mechanism for superconductivity. The isotope 
  experiment, where the hydrogen is replaced with deuterium, was done and the $T_c$ was found to be reduced~\cite{Drozdov2019}. This can be viewed 
  as  a measurement of the total isotope effect, to good approximation, given that the La phonons play little role in $T_c$. If the La had been important, then
   the deuterium substitution could only be viewed as giving a partial isotope effect~\cite{Carbotte:1990}. Simple Bardeen-Cooper-Schrieffer (BCS) theory 
   predicts $T_c\propto M^{-\beta}$, where $M$ is the ion mass for a monatomic crystal and the isotope effect coefficient $\beta=0.5$. However, 
   in Migdal-Eliashberg theory, which is the extension of BCS theory to include the details of the electron-phonon interaction, the isotope coefficient is more complex. It can be reduced significantly from 0.5 with finite $\mu^*$ and even be negative in
   value, as has been seen in experiments on conventional superconductors.     For $\mu^*=0$, $\beta=0.5$ regardless the $\alpha^2F(\omega)$.
   Here, we have calculated the the total isotope effect coefficient to be 0.489 in Eliashberg theory using the functional derivative method of Rainer and Culetto~\cite{Rainer:1979}. Our value is in reasonable agreement with experiment,  which finds 0.46 at 150 GPa~\cite{Drozdov2019}, and with other theoretical calculations showing the pressure dependence of the coefficient~\cite{Durajski:2019}. Our calculation includes the low frequency La-based phonons but even when we remove those phonons to do the partial isotope effect due to hydrogen only, the coefficient barely changes, as expected from the functional derivative.

 \begin{figure} [b!]
 \centering
\includegraphics[width=0.43\textwidth, clip]{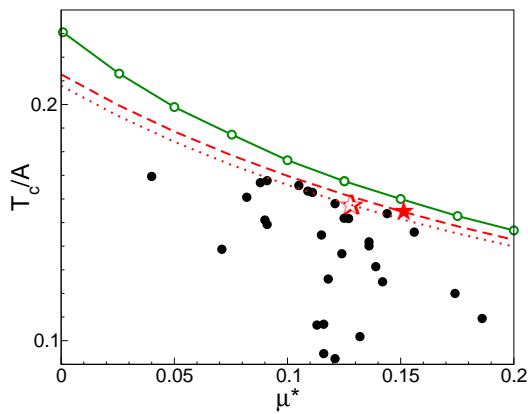}
\caption{\label{Tc_mu} $T_c/A$ versus $\mu^{*}$. The green curve with circles is the upper bound
 allowed by Migdal-Eliashberg theory. The black points are the data points for many conventional 
 superconductors as tabulated in Ref.~\cite{Carbotte:1990} based on primarily $\alpha^2F(\omega)$ spectra measured in 
 experiment via tunneling inversion. The red stars are 
 the values for LaH$_{10}$ using the spectra discussed here. The dashed and dotted curves are calculated 
 with these same two spectra (anharmonic and harmonic, respectively) fixing the area $A$ and varying
  the $\mu^*$. This plot demonstrates that the LaH$_{10}$ spectra are nearly optimal.}
\end{figure}
     
 In Fig.~\ref{A2F}(b), we also show the $\lambda(\omega)$ curve commonly exhibited with $\alpha^2F(\omega)$ spectra in the literature. $\lambda$ is
 the electron-phonon mass renormalization parameter and is calculated using $\lambda=2\int_0^\infty d\omega \alpha^2F(\omega)/\omega$. This
  parameter, which is 2.16 for our spectrum, is very important in the theory and is often taken as a measure of the electron-phonon 
  interaction, appearing in simplified $T_c$ equations, such as the McMillan equation~\cite{McMillan:1968} or the Allen-Dynes equation~\cite{Allen:1975}. A more significant 
  parameter for $T_c$ is the area under the $\alpha^2F(\omega)$, $A=\int_0^\infty d\omega\alpha^2F(\omega)$. 
Shown here is the partial integration of $\lambda$ as a function of energy $\omega$, {\it i.e.}, $\lambda(\omega)=2\int_0^\omega d\Omega\alpha^2F(\Omega)/\Omega$, normalized to $\lambda$.
In comparison, we show the integrated area of the $\alpha^2F(\omega)$ and the $T_c$ values calculated using just the lower
 part of the spectrum up to that energy. One sees that the $T_c$ tracks the area under $\alpha^2F(\omega)$ more closely than the $\lambda$. Indeed,
  it has been discussed in the past that there is an important relationship between the area and $T_c$ with an upper bound on $T_c/A$ provided
   by general Eliashberg theory~\cite{Leavens:1975,Leavens:1974}. This is shown in Fig.~\ref{Tc_mu} where the upper bound of $T_c/A$ is plotted as a function of $\mu^*$ as the green 
   curve with open circles~\cite{Leavens:1975}. Shown are black points 
   from the spectra and data of many conventional superconductors~\cite{Carbotte:1990}. These points always fall below the upper bound. 
    The open and filled red stars are for our harmonic spectrum and the anharmonic Errea spectrum, respectively. The red curves are for
     the fixed $\alpha^2F(\omega)$ spectrum, where we vary $\mu^*$ and therefore the $T_c$ (dashed and dotted curves for anharmonic and
      harmonic curves, respectively). It is clear that both spectra for LaH$_{10}$ are near the upper limit and this means that they are highly optimized in their electron-phonon
      interaction for the highest $T_c$ possible given the repulsive Coulomb potential parameter $\mu^*$.
      Similar results were found for pressurized H$_3$S~\cite{Nicol:2015}.
      Recently, Quan {\it et al.}~\cite{Quan:2019} have also analyzed results for various hydrides and find good agreement with the area correlating with $T_c$. 

 \begin{figure} [t!]
 \centering
\includegraphics[width=0.44\textwidth, clip]{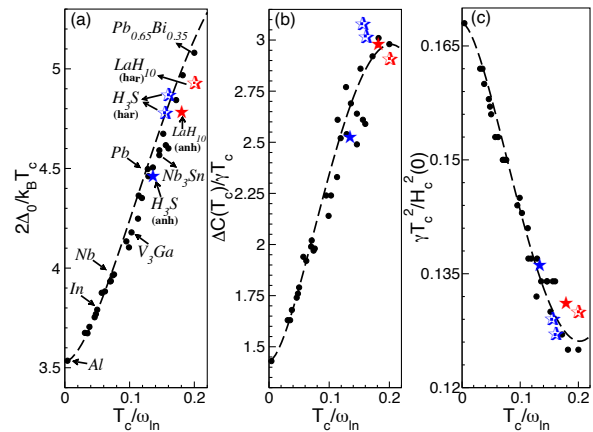}
\caption{The three main BCS ratios as modified by strong electron-phonon 
coupling measured by $T_c/\omega_{\rm ln}$: (a) $2\Delta_0/k_BT_c$, (b) $\Delta C(T_c)/\gamma T_c$, (c) $\gamma T_c/H^2_c(0)$. For $T_c/\omega_{\rm ln}\to 0$, we recover the weak coupling BCS limit of 3.53, 1.43, and 0.168, respectively.
The solid black points are the results for various conventional superconductors 
taken from Ref.~\cite{Carbotte:1990}. The red stars are our 
calculations for LaH$_{10}$ and the blue stars are from previous calculations~\cite{Nicol:2015} for H$_3$S (open and filled stars for the harmonic and anharmonic spectra, respectively). 
  }\label{ratio}
  \end{figure}

\subsection{BCS ratios}

Another important parameter defining the $\alpha^2F(\omega)$ spectrum is a particular moment of the spectrum
 called $\omega_{\rm {ln}}=(2/\lambda)\int_0^\infty d\omega\alpha^2F(\omega) {\rm ln}(\omega)/\omega$~\cite{Allen:1975}. This is treated as a typical  average 
 phonon frequency, replacing the Debye frequency. Over a number of years of intense study of the Eliashberg equations with both numerical 
 and experimental data, the ratio $T_c/\omega_{\rm ln}$ has been identified as one of the best measures of strong electron-phonon coupling 
 in conventional superconductors and simple formulas have been developed using this parameter~\cite{Mitrovic:1984,Marsiglio:1986,Carbotte:1990}. Here, we find $T_c/\omega_{\rm ln}=0.2$ which
  is much stronger coupling than the famous "bad actors" Pb and Hg, and is more typical of Pb-Bi alloys. The Errea spectrum gives a similar value of $T_c/\omega_{\rm ln}=0.18$. The simple formulas for the famous BCS ratios
  which are based on this parameter begin to break down by this value. 
  Indeed, another recent   calculation~\cite{PhysRevB.101.024508} found a spectrum with a value of 0.3. This is quite extreme 
   for conventional superconductors and only amorphous Bi is listed as having this
    value. These authors used the simple formulas for calculating the 
    ratios but in this regime the full numerical theory can give quite significant deviations~\cite{Carbotte:1990}. We have calculated the
     three classic BCS ratios for the energy gap $2\Delta_0/k_BT_c$, the specific heat jump at the phase transition $\Delta C(T_c)/\gamma T_c$ with $\gamma$, the Sommerfeld constant, and the
    thermodynamic critical field $H_c$ at zero temperature $\gamma T_c^2/H^2_c(0)$, using the full Eliashberg equations which includes 
    numerical analytic continuation from the imaginary axis equation results to the real axis in order to determine 
    the energy gap $\Delta_0$. These ratios are listed in 
     Table~1 and we also provide the classic plots of these ratios versus $T_c/\omega_{\rm ln}$, which are shown in Fig.~\ref{ratio}. The dashed curves are the 
     simple analytic formulas which were derived as a functional form from Eliashberg 
    theory and then had the coefficients fitted to the numerical and the experimental data in existence at the time. 
    The fits are best at low $T_c/\omega_{\rm ln}$. We have also included data from many conventional superconductors 
    taken from Ref.~\cite{Carbotte:1990}, some of which are labeled. In the limit of $T_c/\omega_{\rm ln}\to 0$, the 
    BCS values are recovered. Aluminum is a classic weak coupling BCS superconductor and Pb is a well-studied example of a strong
     coupling superconductor. The blue stars are from a prior calculation by one of us~\cite{Nicol:2015} for H$_3$S, where both harmonic~\cite{PhysRevLett.114.157004,Flores-Livas:2016} and anharmonic~\cite{PhysRevLett.114.157004} phonon spectra were used.   The red stars are for 
     our spectrum and the Errea spectrum for LaH$_{10}$. One sees that even at this value of $T_c/\omega_{\rm ln}$ there are
      deviations from the approximate formula, but nonetheless, LaH$_{10}$ is still within the realm 
      of conventional s-wave, electron-phonon-mediated superconductivity. The deviations between the harmonic 
      versus the anharmonic spectrum reflect that the anharmonic one has the hydrogen phonons pushed to
       slightly higher $\omega$ which increases $\omega_{\rm ln}$. 
 \begin{figure} [t!]
 \centering
\includegraphics[width=0.43\textwidth, clip]{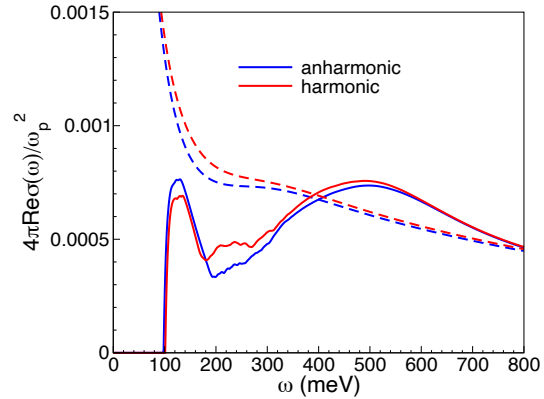}
\caption{\label{conductivity} The real part of the conductivity ${4\pi\rm Re}\sigma(\omega)/\omega^2_P$ as
 a function of photon energy $\omega$. The solid curves are for low temperature $T=0.1T_c$ and the dashed curves are for $T=T_c$.}
\end{figure}
 
 \subsection{Optical Properties}
We now turn to a discussion of optical properties. The reason for focusing on this quantity is twofold. Firstly, the hydrides are
 in diamond anvil cells under high pressure and this limits the type of experiments which can be performed on these new materials in order to verify theories
 about them. It has already been shown that optical experiments can be successfully performed on these materials as was done for H$_3$S~\cite{Capitani2017}.
 Secondly,
 the optical response is a spectroscopy and this can provide essential information about the superconducting state. Indeed at low photon energy,
    one can measure the energy gap and, at higher photon energy, one can
    derive information on the electron-phonon interaction and confirm coupling to high energy bosons.
    Moreover, a recent proposal ~\cite{PhysRevLett.121.047002}
    demonstrates a method
    to provide an independent measure of $T_c$ at high photon energy.

 \begin{figure} [b!]
 \centering
\includegraphics[width=0.43\textwidth, clip]{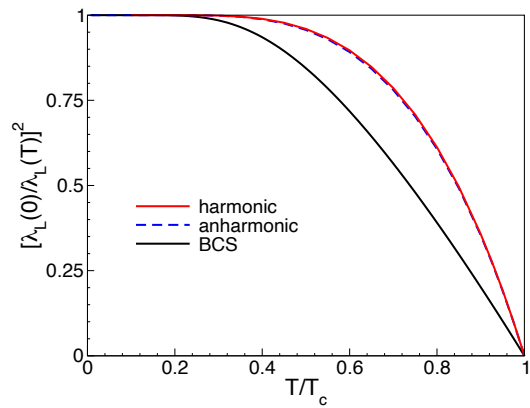}
\caption{\label{Superfluid} Superfluid density or the square of the inverse penetration 
depth $\lambda_L(T)$ versus $T/T_c$. The two LaH$_{10}$ spectra show classic strong coupling effects beyond the BCS curve but otherwise are virtually the same.}
\end{figure}

    In Fig.~\ref{conductivity}, we show the absorptive part of the conductivity ${4\pi\rm Re}\sigma(\omega)/\omega^2_P$ as a function of photon energy $\omega$. Here, $\omega_P$ is the plasma frequency. Curves are shown for the low temperature
    $T/T_c=0.1$ (solid curves) and the normal state at $T_c$ (dashed curves). A comparison has been made between the harmonic and anharmonic spectra, red and blue curves, respectively, and there is very little difference. We have assumed an elastic impurity scattering rate of 100 meV.  The opening of an energy gap in the superconducting state manifests itself as zero absorption up until $2\Delta_0\sim 100$ meV at which point the conductivity rises sharply due to impurity scattering in the system. This region of zero absorption can be used to measure the energy gap~\cite{Mori:2008}. The dip in the absorption reflects structure which can be traced back to energy dependence of the $\alpha^2F(\omega)$ spectrum shifted by $2\Delta_0$. Such structure can be used to extract the $\alpha^2F(\omega)$ ~\cite{Farnworth:1976}. The fact that the main difference between the harmonic and anharmonic spectra occurs in this region illustrates this point as this is the only place where the slightly different energy dependence of the two spectra would manifest.
    The inelastic scattering 
    due to the electron-phonon interaction gives rise to a large bump in the conductivity which overshoots the normal state, primarily reflecting a shift in such structure with the opening an energy gap upon going from the normal to the superconducting state~\cite{Allen:1971,PhysRevLett.121.047002}.
    The stronger variations in the superconducting state over the normal state is a reflection of the square root singularity in the superconducting quasiparticle density of states which enters the evaluation of the conductivity~\cite{Carbotte:2019,PhysRevB.43.12804,Marsiglio:2008}.
    The loss in spectral weight at low energy is found in the superconducting condensate and is represented by the superfluid density shown in Fig.~\ref{conductivity} (related to the inverse square of the penetration depth $\lambda_L(T)$). Here we find that the superfluid density shows the classic temperature dependence of a strong-coupling superconductor where the curve is higher than the BCS one. Both anharmonic and harmonic spectra give essentially the same result here. The penetration depth can be measured from the optical response via the imaginary part of the conductivity for $\omega\to 0$. Note that the imaginary part of $\sigma$ is related to the real part via Kramers-Kronig relations. 
 Finally, returning to the real part of the conductivity in Fig.~\ref{conductivity}, the bump up at high energy here is quite large and its temperature dependence can be used to measure the $T_c$ because the bump will reduce and shift to lower $\omega$ as the energy gap closes with increasing $T$, while the normal state conductivity will increase in width with temperature.

 \begin{figure} [ht!]
 \centering
\includegraphics[width=0.43\textwidth, clip]{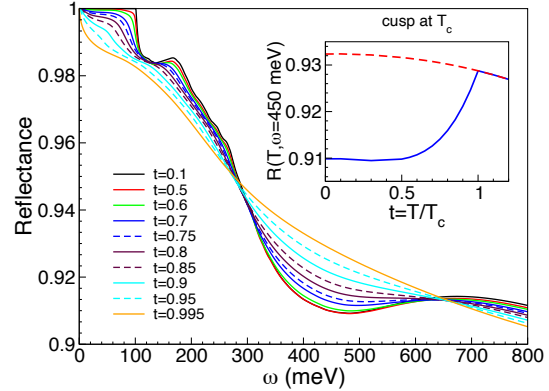}
\caption{\label{Reflectance} Reflectance versus $\omega$ for various temperatures
$t=T/T_c$ in the superconducting state. The inset shows the reflectance as a function of temperature for a 
fixed photon energy of 450 meV. The blue curve below $T_c$ is the superconducting state and the red dashed 
curve is the normal state result that would be measured in the absence of superconductivity.} 
\end{figure}

 To best understand this last point, we move to the reflectance. While both the complex conductivity and the related complex reflectance can be calculated theoretically and measured via experiment, it is the reflectance that is the directly measured quantity. The phase of the reflectance must be obtained through Kramers-Kronig but once this is known, there are simple formulas that provide the connection to the real and imaginary parts of $\sigma$ through the complex dielectric function. In Fig.~\ref{Reflectance}, we show the reflectance that we have calculated for our spectrum at a large number of temperatures from close $0.1T_c$ to $0.995T_c$, using $\omega_P=31$ eV and correcting for the diamond refractive index. The perfect reflection at low temperature and frequencies up to 100 meV identifies the energy gap $2\Delta_0$. The strong variation in the region of 300-600 meV is where the temperature dependence is moving upwards whereas in the normal state it would move downwards. If we fix at a particular photon energy
 (in this case 450 meV) and plot the reflectance as a function of temperature, we have the blue curve shown in the inset, which shows a cusp at $T_c$ as the temperature dependence changes behavior from the superconducting state to the normal state.  The red dashed curve is that for the normal state shown down to zero temperature, if superconductivity is suppressed. This cusp feature in the reflectance at fixed $\omega$ can be used as an additional noncontact method to determine $T_c$~\cite{PhysRevLett.121.047002}. Moreover, as these materials are in a diamond anvil cell in order to be at high pressure, optical measurements are one of the few probes available under these conditions and this range of photon energy is above that where there would be a strong obscuring response due to the diamond in the pressure cell.  This has yet to be tested but successful optical measurements in this energy range were performed on pressurized H$_3$S to confirm coupling to a high energy boson for superconductivity~\cite{Capitani2017} and so we are confident that this experiment is possible and estimates indicate that it should be.

\section{Conclusions}
In summary, we have used a combination of first-principle DFT calculations and Migdal-Eliashberg theory to study the mechanism of superconductivity in LaH$_{10}$, which has the highest known $T_c$. We have demonstrated that the 
large $T_c$ is driven by the coupling of the electrons to hydrogen phonons and that the La contribution to $T_c$ is small. We have provided a thorough discussion of the nature of $T_c$, resulting from a mechanism which is highly optimized in this material, and have shown where LaH$_{10}$ sits in relation to conventional electron-phonon superconductors in terms of strong coupling parameters and the classic BCS ratios.  Finally, 
we have highlighted the optical properties, in particular, as a means to determine the energy gap, the superconducting mechanism, and independently confirm $T_c$. Measurement of optical properties on pressurized H$_3$S has demonstrated the potential for success of this technique and its ability to confirm our predictions.

\section*{Acknowledgments}
S.F.E thanks  Ion Errea, Roald Hoffmann, and Stanimir Bonev for useful discussions. E.J.N. and T.T. acknowledge the late J.P. Carbotte for important discussions and contributions on this topic. In addition, we also thank M.I. Eremets for ongoing collaborations and discussions related to LaH$_{10}$. We also thank H. Liu and R.J. Hemley for providing their $\alpha^2F(\omega)$ from Ref. 25, although it has not been used here. This work was funded by Natural Sciences and Engineering Research Council of Canada (NSERC) (EJN,TT).  This research was enabled in part by support provided by
 Graham \& Cedar and Compute Canada (www.computecanada.ca). \\
\bibliography{ref}

\end{document}